\documentclass[11pt,preprintnumbers,aps,amssymb,nofootinbib,amsmath]
{revtex4}
\usepackage{epsfig,epsf}
\usepackage{graphicx}
\usepackage{verbatim}
\usepackage{epstopdf}
\usepackage{bm} 
\newcommand{\beq}{\begin{equation}}
\newcommand{\beql}[1]{\begin{equation}\label{#1}}
\newcommand{\eeq}{\end{equation}}
\newcommand{\bea}{\begin{eqnarray}}
\newcommand{\eea}{\end{eqnarray}}
\def\eq#1{{(\ref{#1})}}
\def\fig#1{{Fig.~\ref{#1}}}

\newcounter{topiccounter}
\newcommand{\Lb}{\left(}
\newcommand{\Rb}{\right)}

\def\pom{{I\!\!P}}
\def\reg{{I\!\!R}}

\begin{document}

\preprint{TAUP -2916-10\,\,\,
{\tt  [hep-ph]}\,\,
\today\\
}

\title{QCD motivated approach to soft interactions 
at high energy: inclusive production\\}

\author{  E. Gotsman,   E. Levin, \,U. Maor\\}
\affiliation{
Department of Particle Physics, School of Physics and Astronomy\\
Raymond and Beverly Sackler Faculty of Exact Science\\  
Tel Aviv University, Tel Aviv, 69978, Israel\\}

\date{\today}

\pacs{13.85.-t, 13.85.Hd, 11.55.-m, 11.55.Bq}



\begin{abstract}
We extend our two component Pomeron model (GLMM) for soft high energy 
scattering to single inclusive cross sections. 
To this end we present a suitable formulation which also includes 
the semi enhanced Pomeron-particle vertex corrections.
The available data on single inclusive density $(1/\sigma_{in})d\sigma/dy$
in the c.m. energy range of 200-1800 $GeV$ are well reproduced by our model.
The just published ALICE collaboration point at 900 $GeV$ and the CMS 
collaboration measurements at 900 and  2360 $GeV$ are in 
excellent agreement with the calculations of our model .
We also present predictions covering the complete LHC energy range 
which can be readily tested in the early low luminosity LHC runs.
The results presented in this communication
provide additional support to our Pomeron model approach.

\end{abstract}
\maketitle
\par
In this paper we expand our approach to soft hadron interactions, 
developed in Ref.\cite{GLMM}, to obtain estimates for
single inclusive cross sections. 
We have two main objectives: First, we wish to reproduce 
the existing data and predict the single inclusive experimental distributions  
which will be measured in the preliminary low luminosity LHC runs. 
 The data for $\sqrt{s}$ = 200-2360 GeV are well reproduced by our model. 
The first published LHC experimental output\cite{Alice}, 
provides data on p-p single inclusive cross section at $\sqrt{s}$ = 900 GeV 
which is in accord with the results of our model. The recently published
data by the CMS collaboration \cite{CMS} at 900 GeV and the higher energy 
of 
$\sqrt{s}$ = 2.36 TeV  are also in  agreement with our predictions.  
Second, the successful data analysis presented in this paper follows
directly from  our GLMM model and its fitted parameters\cite{GLMM}. 
As such, it provides additional support to the validity of our hypothesis.  
\par
In our approach to soft Pomeron interactions we combine two elements: 
A two channel Good-Walker mechanism\cite{GW} with a super critical Regge-like 
Pomeron with an intercept $\Delta_\pom \,>\,0$, to which we add the 
enhanced (multi) Pomeron interactions. 
Our formulation is based on two main assumptions: 
\newline
1)  We assume that the slope of the Pomeron trajectory
$\alpha^{\prime}_\pom = 0$. This assumption is strongly supported by the 
global data analysis we have presented in Ref.\cite{GLMM}, in which the 
fitted value of $\alpha^{\prime}_\pom$ is exceedingly small. 
A consequence of the small value of  
$\alpha^{\prime}_\pom$ is a relatively large $\Delta_\pom \simeq 0.35$.
\newline
2) In our calculations of enhanced (and semi enhanced) Pomeron interactions 
we only take into account  the triple Pomeron vertex.
\par
These assumptions are compatible with the main features of the Pomeron 
in N=4 super symmetric Yang-Mills theory, in which the Pomeron has an 
intercept of 
$\Delta_\pom = 1 - 2 /\sqrt{\lambda}$ at large values of $\lambda$, 
and $\alpha^{\prime}_\pom\,=\,0$. 
Note that the fitted value $\Delta_\pom \simeq 0.35$ obtained in our 
model corresponds to a large value of $\lambda \simeq 10$. In this approach, 
the main contributions to the total cross section are  the 
elastic and diffractive cross sections. This is a consequence of the  
Good-Walker mechanism\cite{GW} coupled to the vanishing of the cross sections 
initiated by multi Pomeron interactions (for details see Ref.\cite{GLMM}). 
The strength of the  Pomeron interaction is proportional to 
$2/\sqrt{\lambda}$, which can be taken into account by introducing 
a triple Pomeron vertex. 
\par
Our Pomeron model assumptions provide a natural matching between soft Pomeron 
dynamics and high density QCD (hdQCD), 
see Refs.\cite{BFKL,LI,GLR,MUQI,MV,B,K,JIMWLK}. 
The only hdQCD dimensional scale which is responsible for high energy 
interactions is $Q_s$, the saturation scale. This scale 
increases with energy leading to 
$\alpha^{\prime}_\pom \propto 1/Q^2_s (x)\to 0$ at high enough energies where 
$x \rightarrow 0$. 
Recall that the triple BFKL Pomeron vertex plays a decisive  
role in small $x$ pQCD \cite{GLR,B,K,BRN,BART,MUCD,LALE,LELU,LMP}. 
The consequent emerging compatibility of soft and hard Pomeron dynamics 
and their similar formulations are the main results of our Pomeron studies.
\par
\begin{figure}[ht]
\includegraphics[width=100mm]{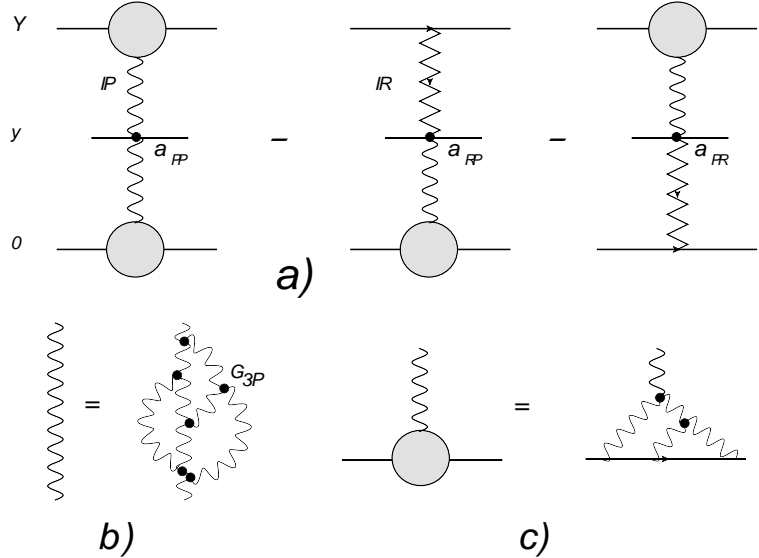}
\caption{ Mueller diagrams\cite{MUDI} for a single inclusive cross section. 
A bold waving line presents the exact Pomeron Green function of \eq{IXS2},  
which is the sum of the enhanced diagrams of \fig{incldi}-b. 
A zig-zag line corresponds to the exchange of a Reggeon.}
\label{incldi}
\end{figure}

In the framework of Pomeron calculus\cite{GRIBRT} 
(see also Refs. \cite{COL,SOFT,LEREG}), single inclusive 
cross sections can be calculated 
using the Mueller diagrams \cite{MUDI} shown in \fig{incldi}-a.  
They lead to 
\bea \label{IXS1}
&&\frac{1}{\sigma_{in}}\frac{d \sigma}{dy}
=\frac{1}{\sigma_{in}(Y)}\left\{a_{PP}(\alpha^2 g_1 
+\beta^2 g_2)^2 G_{enh}\Lb T\Lb Y/2-y\Rb\Rb\times G_{enh}\Lb T\Lb Y/2 
+y\Rb\Rb \right.\\
&&-\left.a_{RP} (\alpha^2 g^R_1 
+\beta^2 g^R_2)(\alpha^2 g_1+\beta^2 g_2)\right. \nonumber\\
&&\left.\left[e^{(\Delta_R (Y/2-y)}\times G_{enh}\Lb T\Lb Y/2+y\Rb\Rb
+e^{(\Delta_R(Y/2-y)}\times G_{enh}\Lb T\Lb Y/2+y\Rb\Rb\right]\right\},
\nonumber
\eea
where the Pomeron Green's function is 
\beq \label{IXS2}
G_{enh}\Lb Y\Rb\,=\,1 - \\exp\Lb \frac{1}{T\Lb Y\Rb}\Rb\,\frac{1}{T\Lb 
Y\Rb}\,\Gamma\Lb 0,\frac{1}{T\Lb Y\Rb} \Rb.
\eeq 
Following Gribov\cite{GRIBRT}, 
we take into account in Eq. (1), the sum of the Pomeron enhanced diagrams, 
considering them as a first approximation for the exact Green function of the 
Pomeron (\fig{incldi}-b). 
Eq. (2) gives the explicit form of this Green function for  
$\alpha^{\prime}_\pom=0$. 
Also included in Eq. (1) are the contributions of the secondary Reggeons. 
\par 
In Eq. (1) we have introduced two new phenomenological parameters, 
$a_{\pom \pom}$ and $a_{\pom \reg}$=$a_{\reg \pom}$, 
for the description of hadron emission  
from the Pomeron and Reggeon. There is
an additional dimensional parameter, denoted by $Q$, which represents the
average transverse momentum of the produced minijets.
$Q_{0}Q$ is the effective mass squared of these minijets, 
with $Q_{0}$ = 2 GeV (see Ref.\cite{KL} for details).
$Q$ and $Q_{0}$ are needed to calculate the pseudorapidity $\eta$
which replaces the rapidity $y$.
The relation between $y$ and $\eta$ is well known 
(see, for example, Ref.\cite{KL}),  
\beq \label{IXS3}
y\Lb \eta, Q\Rb \,\,=
\,\,\frac{1}{2} \ln\left\{\frac{\sqrt{\frac{Q_{0} Q+ Q^2}{Q^2} \,
+\,\sinh^2\eta}\,\,+\,\,\sinh \eta}{\sqrt{\frac{Q_{0} Q+ Q^2}{Q^2}\,
+\,\sinh^2\eta}\,\,-\,\,\sinh \eta}\right\}, 
\eeq
with the Jacobian
\beq \label{IXS4}
h\Lb \eta,Q\Rb \,=\, \frac{\cosh\eta}{\sqrt{\frac{Q_{0} + Q}{Q} 
\,+\,\sinh^2\eta}}.
\eeq
\par
In the parametrization of Ref.\cite{GLMM}, the value of 
the Pomeron-particle vertices are large. To compensate, 
we also sum the semi-enhanced diagrams 
which contribute to the exact vertex of the Pomeron-particle 
interaction (see \fig{incldi}-c). 
This vertex is equal \cite{SCH,BORY} to 
\beq \label{IXS5}
G_{enh}\Lb y\Rb g_i\Lb b\Rb \,\to \,g_i\Lb b, y\Rb\, 
=\,g_i\,G_{enh}(y)\,S_i(b)/(1 + g_iG_{enh}(y)\,S_i(b)),
\eeq
where\cite{GLMM},
\beq\label{IXS6}
S_i(b) = \frac{m^2_i}{4 \pi}\,b\,m_i\,K_1(m _i\,b).
\eeq
Using Eq. (5), we obtain
\bea \label{IXS7}
&&\frac{1}{\sigma_{in}}\,\frac{d \sigma}{d y}\,\,
=\,\,\frac{1}{\sigma_{in}(Y)}\,\left\{ a_{PP}
\left(\int d^2 b (\alpha^2 \,g_1(b,Y/2 - y)
+ \beta^2 g_2(b,Y/2 -y))\right.\right. \nonumber\\
&&\left.\left.\times\,\int d^2 b (\alpha^2 \,g_1(b,Y/2 + y)
+ \beta^2 g_2(b,Y/2 +y)\right)  \right.\\
&&\left.\,\,\,-\,\,a_{RP} \,\,(\alpha^2 \,g^R_1
+ \beta^2 g^R_2)\,(\alpha^2 \,\int d^2 b ( \alpha^2 \,g_1(b, Y/2 - y)
+ \beta^2 g_2(b,Y/2 -y))\,e^{\Delta_R\,(Y/2 + y)}\right.\nonumber\\
&&\left.+\,\int d^2 b ( \alpha^2 \,g_1(b, Y/2 + y)
+ \beta^2 g_2(b,Y/2 +y))\,e^{\Delta_R\,(Y/2 - y)})\right\}.\nonumber
\eea
\begin{figure}[ht]
\includegraphics[width=90mm]{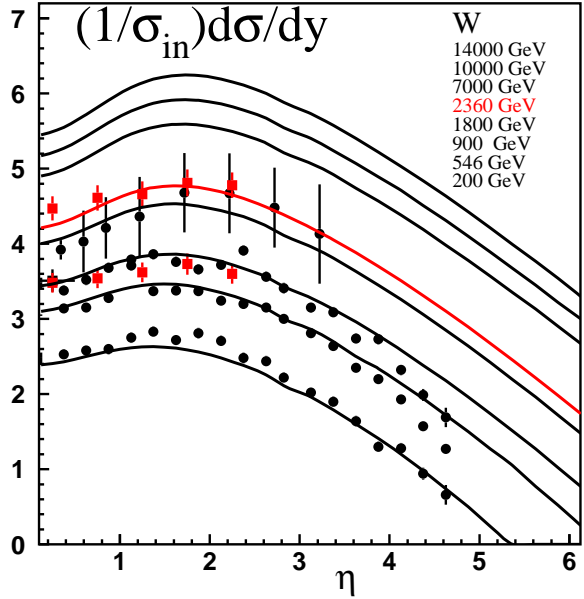}
\caption{Single inclusive density versus energy. 
The dotted data were taken from Ref.\cite{PDG}. 
The square\\ data points correspond to the  
experimental data from LHC by Alice Collaboration\cite{Alice} at W = 900 
GeV and the CMS collaboration\cite{CMS} at W = 900 and 2360 GeV. }
\label{incl}
\end{figure}
Introducing a new notation, 
\beq \label{IXS8}
V(y)\,=\,\int d^2 b  \tilde{V}(b,y)\,\,
=\,\,\int d^2 b ( \alpha^2 \,g_1(b, Y/2 - y) + \beta^2 g_2(b,Y/2 -y)), 
\eeq
we obtain a more compact expression for Eq. (7) 
\bea \label{IXS9}
&&\frac{1}{\sigma_{in}}\,\frac{d \sigma}{d y}\,\,=\,\,\frac{1}
{\sigma_{in}(Y)}\,\left\{ a_{PP}V(y/2-y)V(Y/2+y)\,\right. \nonumber\\
&&\left.-\,a_{RP} \,\,( \alpha^2 \,g^R_1 + \beta^2 g^R_2)\,(V(Y/2 - y)\,
e^{(\Delta_R\,(Y/2 + y)}\,+\,V(Y/2 +y)\,e^{(\Delta_R\,(Y/2 - y)})\right\}. 
\eea 
Eq. (9) enables us to calculate the single inclusive density as a function 
of the pseudo rapidity $\eta$. 
\par
As noted, this calculation entails three additional 
parameters. 
The determination of these parameters from existing data \cite{PDG} is 
not trivial. Comparing the numbers corresponding to the data shown in 
 \fig{incl}, it is evident that a conventional overall 
 $\chi^2$ analysis is impractical, owing to the quoted error bars of the 
 546 GeV data points, which are considerably smaller than the error bars 
quoted for the other energies.  
The full lines in \fig{incl}  are the  results derived from a $\chi^2$ 
fit 
to the  200-1800 GeV data, excluding the 546 GeV points. This fit 
yields a seemingly poor $\chi^2/d.o.f = 3.2$. Despite this, 
we consider this fit to be acceptable, as the data points 
"oscillate" about a uniform  line with error bars
which are much smaller than their deviation from a smooth average.
The results of this fit are
$a_{\pom\pom}$ = 75.7, $a_{\pom\reg}$ = 0.12 and Q = 3.8 GeV.
In our procedure, the line for 546 GeV  in \fig{incl} is calculated
with the model parameters and is visually compatible with the experimental 
data points. Note that both the axes of
 \fig{incl} are linear, 
and that  our calculation 
coincides with the  LHC experimental results \cite{Alice} and \cite{CMS}. 
We have 
also 
made predictions for the higher energies at which the LHC is expected to 
run, see \fig{incl}.      
 The contributions of the secondary Regge trajectories are minimal. 
The experimental values for
$\sigma_{in} = \sigma_{tot} - \sigma_{el} - \sigma_{diff}$
were taken from Refs.\cite{PDG,Alice,CMS}. For our predictions we have 
used
the values of $\sigma_{in}$ calculated in our GLMM model.
Our output over-estimates the few data points with 
$\eta > 4$ data at 546 and 900 GeV by up to 20$\%$. 
This is to be expected, as we have not taken into account the 
parton correlations due to energy conservation, which are important in the 
fragmentation region, but  difficult to include
in the framework of  Pomeron calculus.
\par
To summarize, we have presented a theoretical formulation for single inclusive 
hadron-hadron interactions based on our GLMM model. 
We have reproduced the p-p data\cite{PDG,Alice,CMS} on single inclusive 
density 
as a function of the pseudo rapidity. Our results provide additional support 
for 
our proposed Pomeron approach. 
We have also presented  predictions for the LHC energy range. 
These predictions may soon be tested during the preliminary low luminosity 
LHC runs.\\
\newline
{\bf Acknowledgement:} This research was supported in part by BSF grant  
\#20004019.  

\end{document}